\documentclass[preprint,showpacs,preprintnumbers,floatfix,superscriptaddress,prl,aps]{revtex4}

\usepackage{amssymb}
\usepackage{amsmath}
\usepackage{graphicx}
\usepackage{dcolumn}
\usepackage{bm}

\begin{document}
%\begin{CJK*}{GB}{gbsn}       %% for Yu
%\begin{CJK*}{GBK}{song}       %% for Dong
\title{Plasmoid ejection and secondary current sheet generation from magnetic
reconnection in laser-plasma interaction}

\author{Quan-Li Dong}
\affiliation{Beijing National Laboratory of Condensed Matter Physics, Institute of Physics, Chinese Academy of Sciences, Beijing 100080, China}
\author{Shou-Jun Wang}
\affiliation{Beijing National Laboratory of Condensed Matter Physics, Institute of Physics, Chinese Academy of Sciences, Beijing 100080, China}
\author{Quan-Ming Lu}
\affiliation{CAS Key Laboratory of Basic Plasma Physics, University of Science
and Technology of China, Hefei, Anhui 230026, China}
\author{Can Huang}
\affiliation{CAS Key Laboratory of Basic Plasma Physics, University of Science
and Technology of China, Hefei, Anhui 230026, China}
\author{Da-Wei Yuan}
\affiliation{Beijing National Laboratory of Condensed Matter Physics, Institute of Physics, Chinese Academy of Sciences, Beijing 100080, China}
\author{Xun Liu}
\affiliation{Beijing National Laboratory of Condensed Matter Physics, Institute of Physics, Chinese Academy of Sciences, Beijing 100080, China}
\author{Yu-Tong Li}
\affiliation{Beijing National Laboratory of Condensed Matter Physics, Institute of Physics, Chinese Academy of Sciences, Beijing 100080, China}
\author{Xiao-Xuan Lin}
\affiliation{Beijing National Laboratory of Condensed Matter Physics, Institute of Physics, Chinese Academy of Sciences, Beijing 100080, China}
\author{Hui-Gang Wei}
\affiliation{National Astronomical Observatories of China, Chinese Academy of
Sciences. Beijing 100012, China}
\author{Jia-Yong Zhong}
\affiliation{National Astronomical Observatories of China, Chinese Academy of
Sciences. Beijing 100012, China}
\author{Jian-Rong Shi}
\affiliation{National Astronomical Observatories of China, Chinese Academy of
Sciences. Beijing 100012, China}
\author{Shao-En Jiang}
\affiliation{Research Center for Laser Fusion, China Academy of Engineering Physics, Mianyang 621900, China}
\author{Yong-Kun Ding}
\affiliation{Research Center for Laser Fusion, China Academy of Engineering
Physics, Mianyang 621900, China}
\author{Bo-Bin Jiang}
\affiliation{Research Center for Laser Fusion, China Academy of Engineering
Physics, Mianyang 621900, China}
\author{Kai Du}
\affiliation{Research Center for Laser Fusion, China Academy of Engineering
Physics, Mianyang 621900, China}
\author{Xian-Tu He}
\affiliation{Institute of Applied Physics and Computational Mathematics,
Beijing 100094, China} \affiliation{Institute for Fusion Theory and Simulation, Physics Department, Zhejiang University, Hangzhou 310027, China}
\author{M. Y. Yu}
\affiliation{Institute for Fusion Theory and Simulation, Physics Department,
Zhejiang University, Hangzhou 310027, China} \affiliation{Institute for
Theoretical Physics I, Ruhr University, D-44780 Bochum, Germany}
\author{C. S. Liu}
\affiliation{Department of Physics, University of Maryland, College Park,
Maryland 20742, USA}
\author{Shui Wang}
\affiliation{CAS Key Laboratory of Basic Plasma Physics, University of Science
and Technology of China, Hefei, Anhui 230026, China}
\author{Yong-Jian Tang}
\affiliation{Research Center for Laser Fusion, China Academy of Engineering Physics, Mianyang 621900, China}
\author{Jian-Qiang Zhu}
\affiliation{National Laboratory on High Power Lasers and Physics, Shanghai, 201800, China}
\author{Gang Zhao}
\affiliation{National Astronomical Observatories of China, Chinese Academy of
Sciences, Beijing 100012, China}
\author{Zheng-Ming Sheng}
\affiliation{Beijing National Laboratory of Condensed Matter Physics, Institute of Physics, Chinese Academy of Sciences, Beijing 100080, China}
\affiliation{Department of Physics, Shanghai Jiaotong University, Shanghai
200240, China}
\author{Jie Zhang}
\affiliation{Beijing National Laboratory of Condensed Matter Physics, Institute of Physics, Chinese Academy of Sciences, Beijing 100080, China}
\affiliation{Department of Physics, Shanghai Jiaotong University, Shanghai
200240, China}

\date{\today}

\begin{abstract}
Reconnection of the self-generated magnetic fields in laser-plasma interaction was first investigated experimentally by Nilson {\it et al.} [Phys. Rev. Lett. {\bf 97}, 255001 (2006)] by shining two laser pulses a distance apart on a solid target layer. An elongated current sheet (CS) was observed in the plasma between the two laser spots. In order to more closely model magnetotail reconnection, here two side-by-side thin target layers, instead of a single one, are used.  It is found that at one end of the elongated CS a fan-like electron outflow region including three well-collimated electron jets appears. The ($>1$ MeV) tail of the jet energy distribution exhibits a power-law scaling. The enhanced electron acceleration is attributed to the intense inductive electric field in the narrow electron dominated reconnection region, as well as additional acceleration as they are trapped inside the rapidly moving plasmoid formed in and ejected from the CS. The ejection also induces a secondary CS.
\end{abstract}

%% and Li {\it et al.} [Phys. Rev. Lett. {\bf 99}, 055001 (2007)]

\pacs{52.38.Fz, 52.35.Vd, 52.30.-q, 52.50.Jm}

\maketitle
%\end{CJK*}

Magnetic reconnection (MR) in plasmas is associated with explosive conversion of magnetic energy into plasma kinetic and thermal energies. In the process, plasma is accelerated in and ejected from a thin region where reconnection takes place \cite{Parker1957,Sweet1958,MasudaInnes1990s,Birn2007,Yamada2010}.
The reconnection rate obtained from the experiments and solar
observations is often found to be larger than that predicted by
the standard Sweet-Parker and related models
\cite{Ji2004,Birn2007,Yamada2010} and has been attributed to
various collisionless and higher-dimensional effects, including
Hall current and turbulence \cite{MBSCMR,Zhong2010,Khotya10,Willingale2010,Fox2011,LapentaDaughton2011}.
It has also been shown that the reconnection rate is enhanced by
the formation of secondary magnetic islands and plasmoid ejection from the region, which can be highly unstable if the Lundquist number
$S>10^4$ \cite{Biskamp86DaughtonSamtaney09}. These theoretical
predictions are consistent with {\it in situ} observation of a
secondary magnetic island near the center of the ion diffusion
region in the near Earth magnetotail \cite{Wang2010}. It is well
known that in laser-matter interaction megagauss magnetic fields
can be generated through the barotropic ($\nabla n_e\times \nabla
T_e$) mechanism \cite{Stamper1971,Borg98Petra09}. Nilson {\it et
al.} \cite{Nilson06} were the first to use two approaching laser-created plasmas to model the reconnection geometry. Definitive proof of MR taking place is provided by the measurements of Nilson {\it et al.} \cite{Nilson06} and Li {\it et al.} \cite{LiCK07} on magnetic topology changes using
time-resolved proton deflection technique, and by the observations
of Nilson {\it et al.} \cite{Nilson06} on the highly collimated bi-directional plasma jets $40^\circ$ off the expected MR plane. In this Letter, we investigate the structure of collisionless reconnection of the self-generated magnetic fields generated in laser-target interaction
produced plasmas by using two coplanar plane targets with a gap in between, so that the colliding magnetic fields and plasmas are not electrically pre-connected. Three well-collimated energetic (MeV) electron jets, or elongated electron diffusion regions (EDRs), inside and along the edges of a fan-like plasma outflow from the thin current sheet (CS) are observed. The energy gain is attributed to electron acceleration by the intense local inductive electric fields as well as inside a fast propagating plasmoid ejected from the elongated CS during the MR. The plasmoid also modifies the local magnetic field as it moves forward, producing a secondary CS. Our results agree quite well with that from numerical simulations and in situ observations.
%%Three elongated electron-diffusion regions or
%% to serve respectively as the
%%bright edges and the bright core of the two-dimensional fan-like plasma outflow
%% of the primary CS. The well-collimated electrons are measured to
%% have energy of MeVs,

The experiments were performed at the SG-II laser facility. The
general setup is similar to that in our earlier work
\cite{Zhong2010}. Here, two 0.7 mm $\times$ 0.3 mm $\times$ 50 $\mu$m
thin Al foils separated by 150-240 $\mu$m are juxtaposed in the
same (vertical) plane. Two heater laser beams are simultaneously
(within $\pm 25$ ps) focused by $F=3$ optics onto each foil. The
focal spots, $400\pm25 \mu$m apart, have diameters $<150 \mu$m
FWHM. The incident laser intensity at each spot is $\sim {\cal
O}(10^{15})$ Wcm$^{-2}$, and the total laser energy on each target
is $\sim 450$ joules. The main diagnostics for the density of the
laser-produced plasma is a modified Nomarski interferometer of
magnification 3.5. A 527 nm 150 ps FWHM Gaussian probe laser beam
is directed perpendicular to the target plane at suitable intervals with respect to the heater beams. A polarization imaging system is used to measure the self-emission profile of the plasma, and a combination of reflective and bandpass ($\lambda\sim532\pm1$ nm) interference filters is used to limit the detectable emissions. Three pinhole cameras with magnification 10 are used to
monitor the x-ray emission from the front, back, and side of the
plasma. At the top front of the target plane, an adjustable-range
RAP crystal spectrometer of $\sim$60 $\mu$m spatial resolution is
used to characterize the plasma composition.

\begin{figure}[tbp]\suppressfloats
%\graph2
%%\begin{center}
%%%%\includegraphics*[bb=143 316 442 691, width=10cm,clip]{Fig1.eps}
\includegraphics[width=8.5cm]{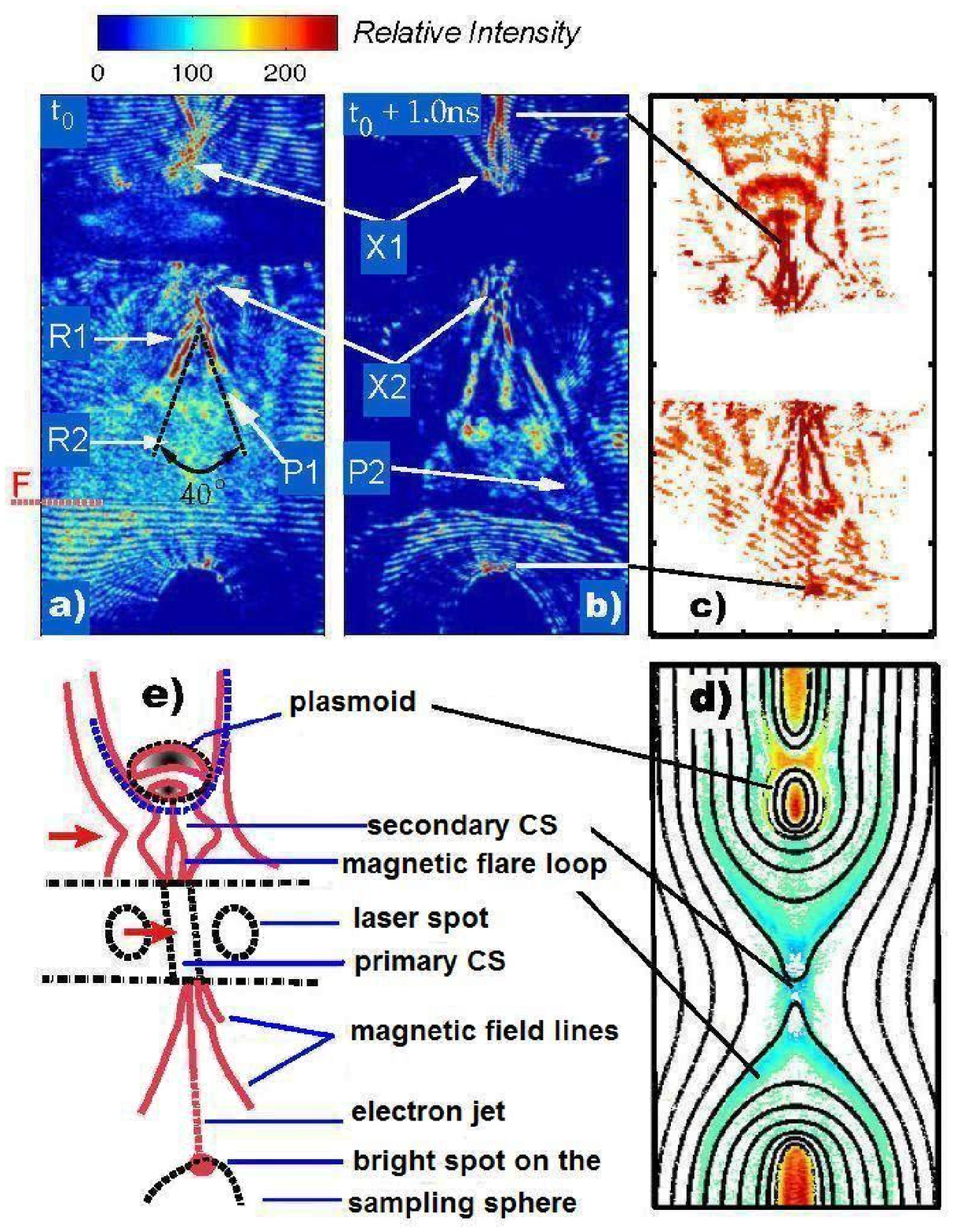}%%{Fig2.jpg}
%%\end{center}
\caption{(Color) Interferometric image at $t=t_0$ ns (a) and
$t=t_0+1$ ns (b), where $t_0$ is set at the half maximum of the
laser-pulse tail. The electron diffusion region is between X1 and
X2. F marks the edge of the advecting plasma. P1 and P2 (1.16 mm
and 1.75 mm from X2, respectively) mark the location of a front of
the expanding fan-like region at the two times, from which an
outflow speed of $>600$ km/s is estimated. The angle of the
fan-like plasma region is 35$^{\circ}$ -- 40$^{\circ}$. (c)
Linearly polarized self-emission image of the plasma. (d) PIC
simulation result showing the current density, typical field
lines, the secondary CS, and a plasmoid. (e) Schematic
illustration of (c). The short thick (red) arrows indicate the
plasma inflow directions for the primary and secondary
CS.\label{f1}}

\end{figure}

\begin{figure}[tbp]\suppressfloats
%\graph3
\begin{center}
\includegraphics[width=8.5cm]{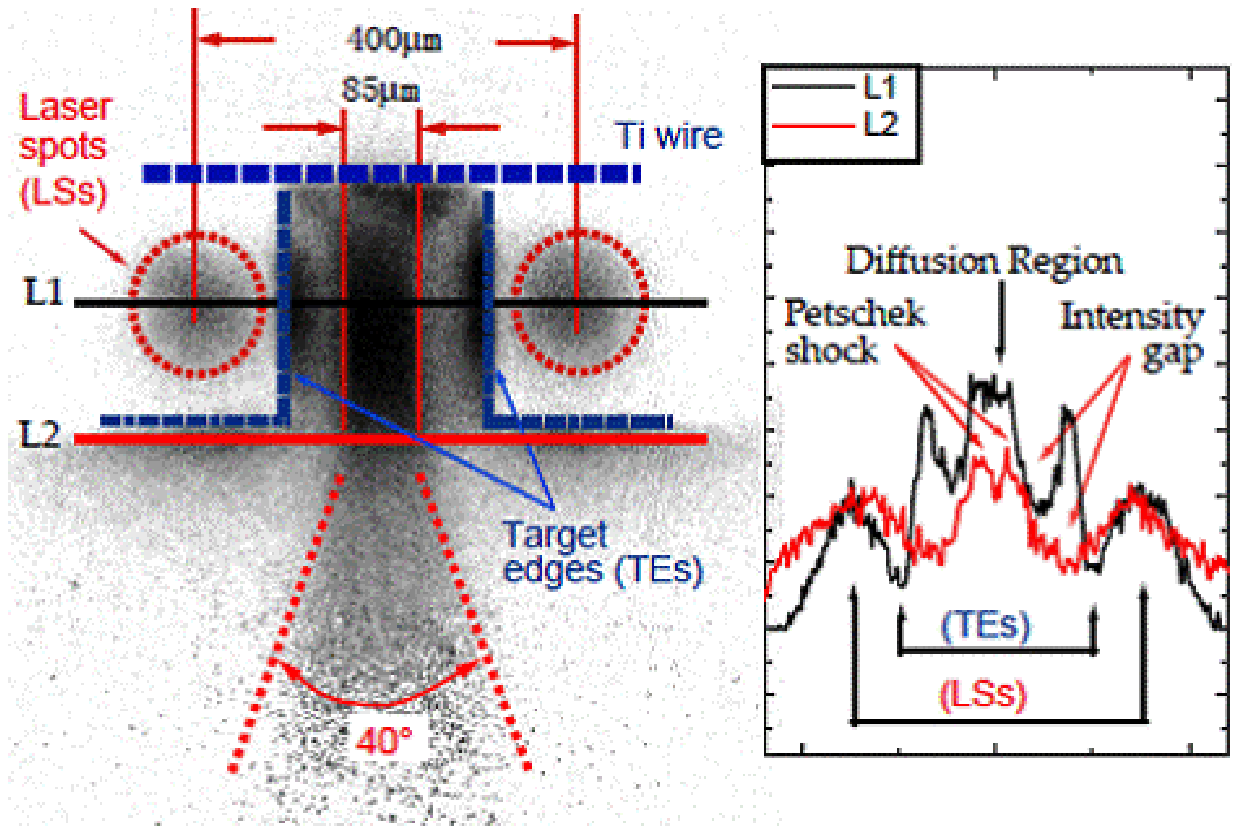}%%{fig3.eps}
\end{center}
\caption{A typical x-ray pinhole image, with superimposed schematics
of the experimental setup, and the corresponding intensity curves
at two locations. The x-ray emission is strongest in the electron
diffusion region at the center, and is of width $\sim85 \mu$m. The two central emission peaks
in the L2 profile correspond to the two edge electron jets
at the start of the fan-like region. \label{f2}}
\end{figure}

Figures \ref{f1}(a) and (b) show the interferometric images of the
laser-generated plasma at $t=t_0$ and $t=t_0+1.0$ ns,
respectively, where $t=t_0$ corresponds to the instant when the
tail-side half-maximum of the heater pulse reaches the target
front. By symmetry the MR region is centered in the gap between
the laser spots \cite{Nilson06,LiCK07}. Figure \ref{f1}(c) shows
the profile of linearly polarized 532 nm emission from the plasma,
with the darkest shade representing the brightest emission. Fig.\
\ref{f1}(d) shows the {\it upper half} of the MR region as
obtained from a particle-in-cell (PIC) simulation using the
experimental parameters (details given below).
% The solid lines between Figs.\ \ref{f1}(b) and 1(c) indicate their positions relative to each other. \
The main features in Fig.\ \ref{f1}(c) are sketched and identified
in Fig.\ \ref{f1}(e), where the solid (red) lines show the
brightest emission regions. All the panels in Fig.\ \ref{f1} and
the left panel in Fig.\ \ref{f2} show the MR, or the $x-z$ plane,
where $x$ is along the two laser spots. By following an edge,
marked F in Fig.\ \ref{f1}(a), the local plasma advection velocity
is determined to be $\sim 2000$ km/s \cite{Willingale2010}.  The
CS region (from X1 to X2) between the two laser spots is of length
$\sim 350 \mu$m. Here are two characteristic regions: R1 points to
a fan-like (with full-opening angle $2\theta_{\rm rec}<
40^{\circ}$) region of irregular plasma motion, and R2 points to a
relatively calm and uniform plasma region. Fig.\ \ref{f1}(b) shows
that after $1.0$ ns the fan-like region has expanded and the jets
lengthened. In the emission image in Fig.\ \ref{f1}(c) one can
clearly see the two bright (solid) lines marking the edges of the
magnetic-separatrix region (see also Fig.\ \ref{f1}(e)). At the
center of the fan-like region, a third bright line corresponding
to a well-collimated energetic-electron jet appears. The three
electron jets have similar widths ($\sim 55\mu$m) in the optical
images. The variance of the lengthes of the three jets in the
time-resolved interferometric images indicates that the center jet
is much faster than that on the two edges of the outflow fan, which has a bulk
velocity $v_{cf}>600$ km/s, and that the center part of the CS has
lengthened. Similar jet-like electron outflows have been observed
in the magnetosheath and the corresponding numerical simulations
\cite{ShayPhanKarim2007}, as well as around the separatrices of
the reconnection layer \cite{Mozer2005}.

The upper half of the MR region in Fig.\ \ref{f1} involves more complex behavior. There is also a fan-like outflow of plasma from the CS. However, away from the primary CS there is a secondary CS with local MR. The secondary CS is connected to a flare loop at one end and a plasmoid at the other end, which in
Fig.\ \ref{f1}(e) is encircled by a thick black dashed line.
Within the latter one can see two bright areas (marked by red
solid lines) that correspond to the leading and trailing ends of
the plasmoid. One can also see that the trailing end of the
plasmoid is connected to the secondary current sheet. A plasmoid
can also be seen in the simulation result shown in Fig.\
\ref{f1}(d). It is born inside the primary CS and grows rapidly.
Its rapid ejection stretches the reconnected magnetic field lines,
creating a secondary CS and MR. In Fig.\ \ref{f1}(c) of the
experimental results, the plasmoid's upward-transport is
manifested by its motion-blurred emission images. However, the
generation process of the plasmoid needs further more detailed
studies. These identifiable features in Fig.\ \ref{f1}(c) are also
marked in Fig.\ \ref{f1}(e) and shall be discussed in more
detail later. Our results thus support that of Shepherd {\it et al.} \cite{Shepherd2010} and Uzdensky {\it et al.} \cite{Uzdensky2010}, and should be useful for analyzing abrupt solar and astrophysical events \cite{LinShimizu,Uzdensky2011}. The top-bottom asymmetry of the MR region is
inevitable in the case with plasmoid ejection
because the CS is highly unstable. Any small local
disturbance can lead to rapid plasmoid ejection. That is, the ejection is stochastic, as proposed by Uzdensky {\it et al.} \cite{Uzdensky2010}.

\begin{figure}[tbp]\suppressfloats
%\graph4
%\begin{center}
%%%\includegraphics*[bb=158 462 436 642, width=10cm,clip]{Fig3.eps}
\includegraphics[width=8.5cm]{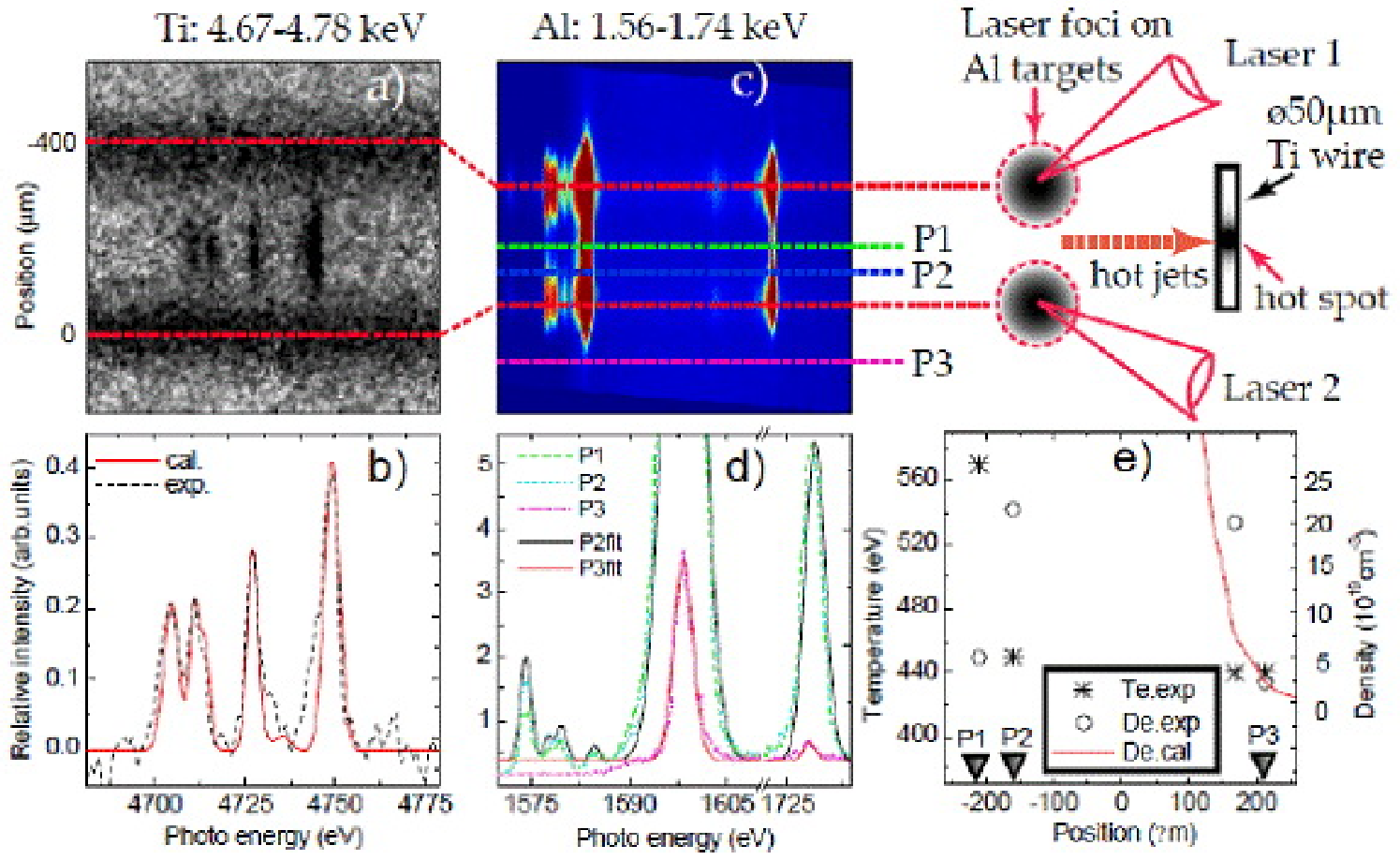}%%{Fig4.pdf}
%%\end{center}
\caption{(Color) Spatially resolved emission of $K_\alpha$ lines
from helium-like ions and the satellite lines from (a) the Ti
wire, and (c) the Al targets. (b) and (d) Typical line spectra
from Ti and Al, respectively, together with the results calculated
using a non-LTE model. (e) Local Al plasma electron density and
temperature estimated from (d). The solid and dashed lines show
the plasma electron density and temperature, respectively,
obtained from a hydrodynamic code. The sketch shows the relevant
experimental components.  \label{f3}}
\end{figure}

More detailed characteristics of the primary CS are obtained from
the x-ray images and the emission spectra.  Fig.\ \ref{f2} shows the
x-ray pinhole image taken from the back of the Al targets. Intense
x-ray emission comes from a narrow region of hot and dense plasma
between the laser spots, with less intense x-ray emission comes
from a fan-like region below it. The former region, of width $\sim
85 \mu$m, can be taken to be the ion diffusion region. A Ti wire
of diameter $50 \mu$m placed $\sim 250 \mu$m above and parallel to
the target plane is used to determine the extent of the primary CS
normal to the MR plane. As shown in Fig.\ \ref{f2}, plasma
ejection appears on the lower side without the Ti wire, but that
on the upper side is mostly blocked by the latter. The extent of
the CS is determined to be about 50 $\mu$m, indicating a quasi
two-dimensional (2D) nature of the MR region
\cite{ShayPhanKarim2007}. However, at longer times the plasma
system becomes highly irregular and difficult to follow with our
diagnostics. This may be due to onset of transverse instabilities
that lead to a strongly turbulent state
\cite{LapentaDaughton2011}. The spatial distributions of the
$K_\alpha$ resonance lines (He-$\alpha$ lines) of the helium-like
Ti ions and their satellites, as shown in Fig.\ \ref{f3}(a), are
used to determine the cross-sectional size of the energetic
electron outflow \cite{Oka2008}. Fig.\ \ref{f3}(b) shows the
spatially integrated line spectra. Although the Ti wire is also in
the transverse expansion paths of the two laser-produced Al
plasmas, the He$-\alpha$ lines of Ti plasmas are only present in
the midplane region (between $80-150 \mu$m). Accordingly, together
with the evidence given by the bright spot created by the central
jet on the sampling sphere, as can be seen in the lower parts of
Figs.\ \ref{f1}(b), (c), and (e), one can conclude that a
well-collimated energetic electron jet is formed. The uniform
continuum emission on both sides of the Ti He-$\alpha$ lines is
from the Al plasma. Simulations by Shay {\it et al.} suggest that
the structure of the fan-like MR plasma outflow and
well-collimated energetic electron jets can be attributed to
competition and balance of the inductive electric field, the
electron inertia, the Lorentz force, and divergence of the
momentum flux \cite{ShayPhanKarim2007}. However, more comprehensive
laboratory and {\it in situ} magnetotail measurements are still
needed in order to more definitively understand the detailed physics of
the process \cite{Mozer2005,ShayPhanKarim2007}.

Figure \ref{f3}(c) shows simultaneously the measured He-$\alpha$ and
Ly-$\alpha$ line emissions of the Al plasma, and Figs.\ \ref{f3}(d) shows the space-resolved line spectra. The x-ray spectra of
the Al plasma at the free-expansion edge, namely along P3 in Fig.\
\ref{f3}(c), shows a clear He-$\alpha$ line and a weak Ly-$\alpha$
line. From non-LTE calculations, we found that this plasma region
has temperature $\sim$ 440 eV and density $\sim 2.5\times10^{19}
$cm$^{-3}$, which are consistent with the interferometric
measurements. The x-ray spectra from the plasma at the center,
namely along P1 in Fig.\ \ref{f3}(c), clearly shows a saturated
He-$\alpha$ line and a strong Ly-$\alpha$ line, as well as several
satellite lines from Li-like ions, indicating an electron
temperature of $\sim570$ eV and density of $\sim 4.8\times
10^{19}$cm$^{-3}$, the latter about double that at the
free-expansion edge (i.e., along P3). Fig.\ \ref{f3}(e) shows the
electron temperature and density from the experiments and
hydrodynamic calculations. However, the Ti plasma produced by
impact of the energetic jet electrons with the Ti wire has a
temperature of $\sim$2300$\pm100$ eV and density of
3.2$\pm$0.5$\times$10$^{23}$ cm$^{-3}$, which are both
considerably higher than that in the diffusion region. This is due
to the fact that the collimated jet electrons have speeds much
higher than the speed $\sim 600$ km/s obtained from the intense
linearly polarized 532 nm synchrotron radiation imaging of the
plasma profile discussed above.

The electron and ion inertial lengths as calculated are
$d_e=c/\omega_{pe}$$=0.77 - 1.06\mu$m and $d_i=c/\omega_{pi}\sim
54 - 75\mu$m, respectively, where $c$ is the light speed, and
$\omega_{pe}$($\omega_{pi}$) is the electron(ion) plasma
frequency. The electron density is $n_e\sim 2.5-4.8\times 10^{19}$
cm$^{-3}$, and the average ionic charge is $Z\sim 10$.
Accordingly, the width $\sim 85 \mu$m of the defined ion diffusion
region in Fig.\ \ref{f2} is $\sim 1 - 1.5$ times the ion inertial
length. The widths of the EDRs are obtained by two methods, which are sensitive to the photon energy ranges. The pinhole x-ray image at $>$1.5keV gives the widths of the EDRs as 7 and 15 $\mu$m as in Fig.2, corresponding
to $\sim 10-15d_e$, respectively. On the other hand, the polarized
emission image at 532 nm gives the widths of the three EDRs as
$\sim 55\mu$m. The difference in the EDR widths from the two
measurement methods indicates that, as expected, the center
regions of the EDRs are of higher temperature. The electron
mean-free-path is calculated as $\lambda_{emfp}\sim 114\mu$m,
which is larger than the measured current layer width.  For the
experimental parameters, we have the ratio between the ion skin
depth and the Sweet-Parker width $d_i/d_{sp}\sim 10>1$, so that
the Hall effect should be the main effect for the fast
reconnection \cite{Yamada2006}, and the binary collision is thus
negligible in the CS, as also assumed for our PIC simulations.
With detailed characteristics of the MR plasma region, and the
separatrix magnetic field of $B_0\sim 3.75$ MG inferred later, we
now can the give the upper limit of the Alfv\'en speed as $V_A\sim
720$ km/s. The measured velocity $\sim$600km/s of the two side
EDRs in Fig.\ \ref{f1} is near the calculated value of $V_A$.
However, the central EDR that starts later than the two side ones
has clearly a higher propagation speed (since it can almost catch
up with the latter), perhaps super-Alfv\'enic as predicted by PIC
simulations.

We have carried out 2D PIC simulations of the main MR region using
parameters close to that of our experiments. However, a large
($m_i/m_e$=100) electron mass and a relatively low ($c/V_A=30$)
light speed are used \cite{Fu2006}. The initial state is a Harris
current sheet of width $d_i$ in a $L_x\times L_z=12.8d_i\times
102.4d_i$ simulation box. A guide field $B_y\sim 0.2B_0$, where
$B_0 {\hat{z}}$ is the strength of the sheared magnetic field, is
added to model the experimental magnetic topology. Initially, the
plasma temperature is $T_{i0}=T_{e0}=0.011m_ec^2$, and the plasma
beta is $\beta_0=0.2$. Fig.\ \ref{f1}(d) shows a simulated current
density distribution $j_y$ (color coded) in the {\it upper half}
of the MR region, together with several typical field lines (black
lines). Here one also finds a plasmoid, together with the deformed
magnetic field lines and a secondary CS, as observed in the
experiments. However, it should be noted that our simulation does
not include the initial laser-plasma interaction stage since that
would be beyond our computational resources. In fact, it only
models the thin central MR region, which should be qualitatively
similar for most MRs \cite{MBSCMR}, and the initial values of the
plasma and field parameters have also been taken from the already
evolving system in the experiment.

The energetic jet electrons from the diffusion region are also
diagnosed by using a magnetic spectrometer \cite{Zhang2005}. The
jet electron energy can reach MeVs and obeys a power-law (of index
$-6.5$) scaling at the high-energy tail for electrons up to
$\sim2$ MeV (the limit of our measuring equipment). We have also
estimated the energy of the relativistic electrons from the
synchrotron radiation (see Fig.\ \ref{f1}(c)). For relativistic
electrons gyrating in the MR and Hall magnetic fields, the maximum
electron energy ${\cal E}_{\max}$ is related to the maximum
measured wavelength $\lambda_{\max}$ of the linearly polarized
synchrotron radiation by ${\cal E}_{\max}= m_ec^2(2.5\times10^{-2}/\lambda_{\max}B_0)^{1/2}$, and
the corresponding electron gyroradius is
$r_g=1.70\times10^3\sqrt{({\cal E}_{\max}/0.511)^2-1.0}/B_0$.
Assuming that the width of the electron diffusion region is of the
order of $r_g\sim55\mu$m,and the maximum wavelength is
$\lambda_{\max}=532$nm, we obtain 5.2 MeV as the measured energy
of the relativistic electrons and $B_0\sim375$ Tesla as the
strength of the reconnected magnetic field. Existing numerical
simulation and analytical estimate indicate that electrons can be
accelerated to ultrahigh energies by the electric fields produced
in MR, which explains the good collimation of energetic electrons
\cite{Drake2005,Drake2006,Pritchett2006,Fu2006}. For electrons to
achieve MeV energy in $\sim100$ ps (the tolerant time for Fig.\
\ref{f1}(c)) to prevent from being blurred by plasma expansion;
also see \cite{Nilson06,LiCK07}), the electric field strength
should be $\sim10^8$ V/m. According to Faraday's law, such a
reconnection electric field corresponds to the annihilation of
several tens Tesla of magnetic fields in $\sim 0.1$ ns, or a
dimensionless reconnection rate \cite{ShayPhanKarim2007} $0.1$ --
$0.3$, which is consistent with the Hall MR rate
$\sim\tan(\theta_{\rm rec})\sim0.25$ -- $0.35$ obtained from
measuring the local magnetic exhaust wedge angle $2\theta_{\rm
rec}$ in Fig.\ \ref{f1}. The corresponding acceleration length
should then be $\sim1$cm. That is, the pre-accelerated electrons
trapped by the rapidly moving plasmoid have bounced many times
before they gain sufficient energy to escape \cite{Wang2010accl}).
Accordingly, the MR electric field and plasmoid trapping,
accounting for the good collimation and the power-law scaling,
respectively, are both responsible for the observed
characteristics of the high-energy electrons.

In summary, two suitably juxtaposed Al foil targets are used to
produce in the laser generated plasma the configuration of MR in
the presence of a guide field. It is found that in the MR a
plasmoid is generated and ejected from the CS. As it rapidly
propagates away it deforms the reconnected magnetic field and
generates a secondary CS as well as flare loops. The primary MR
also involves fan-like plasma outflows, in which appears intense
well-collimated electron jets containing MeV electrons. The
results are consistent with the theoretical prediction of
anomalous plasmoid generation by unstable CS, often invoked to
interpret solar flares. Simulations show that the guide field can
indeed lead to the fast formation of the plasmoid and the
secondary CS. The results here also demonstrate that suitably
designed laser-plasma interactions can be effective and economical
for laboratory modeling high-energy-density astronomical and
astrophysical phenomena.

\begin{acknowledgments}\suppressfloats
This work is jointly supported by the National Natural Science
Foundation of China (grant nos. 10835003, 11074297), the CAS
project KJCX2-YWT01, and the National Basic Research Program of
China (grant nos. 2007CB815101, 2008CB717806). We also thank the
staff of the target manufacture group at Research Center for Laser
Fusion, and the staff of the National Laboratory on High Power
Lasers and Physics for their help in the experiments.
\end{acknowledgments}

%\pagebreak
%% Borghesi1998, LiCK2006, Nilson2006, LiCK2007, Petrasso2009

%\newpage

%\newpage

%\newpage

\end{document}